\documentclass[prb,aps,twocolumn,superscriptaddress]{revtex4-1}
\usepackage{graphicx,color}
\usepackage{amsthm}
\usepackage{amsfonts}
\usepackage{algorithmic}
\usepackage{enumerate}
\usepackage{latexsym}
\usepackage{amsmath}
\usepackage{amssymb}
\usepackage[colorlinks=true,citecolor=blue,linkcolor=blue]{hyperref}

\emergencystretch=\maxdimen
\begin{document}

\title{Doping-dependent metal-insulator transition in a disordered Hubbard model}

\author{Lingyu Tian}
\affiliation{Department of Physics, Beijing Normal University, Beijing
100875, China}
\author{Yueqi Li}
\email{L. Tian and Y. Li contributed equally to this work.}
\affiliation{Department of Physics, Beijing Normal University, Beijing
100875, China}
\author{Ying Liang}
\affiliation{Department of Physics, Beijing Normal University, Beijing
100875, China}
\author{Tianxing Ma}
\email{txma@bnu.edu.cn}
\affiliation{Department of Physics, Beijing Normal University, Beijing
100875, China}
\affiliation{Beijing Computational Science Research Center, Beijing
100193, China}

\begin{abstract}
We study the effect of disorder and doping on the metal-insulator transition in a repulsive Hubbard model on a square lattice using the determinant quantum Monte Carlo method. First, with the aim of making our results reliable, we compute the sign problem with various parameters such as temperature, disorder, on-site interactions, and lattice size.
We show that in the presence of randomness in the hopping elements, the metal-insulator transition occurs and the critical disorder strength differs at different fillings.
We also demonstrate that doping is a driving force behind the metal-insulator transition.
\end{abstract}
\maketitle

\section{Introduction}
\label{sec:intro}
Metal-insulator transitions are an interesting topic of intense activity in modern physics.
In general, there are three kinds of insulators.
Systems in which the valence band is completely filled are called band insulators\cite{PhysRevLett.83.2014,https://doi.org/10.1002/andp.19283921704}.
The staggered potential, in which the on-site energies are different, can also produce band insulators with a spectral gap in cold atom experiments\cite{PhysRevLett.115.115303,PhysRevLett.119.230403}.
In real materials, disorder weakens the constructive interference and affects quantum transport.
Disorder-induced localization, which was proposed more than half a century ago as the Anderson insulator,
has inspired numerous efforts to explore the metal-insulator transition\cite{PhysRev.109.1492,RevModPhys.50.191,PhysRevLett.96.063904}.
In addition to these systems, when electron correlations are considered, a metallic system can become an insulator, induced by the competition between the energy gap and the kinetic energy;
the electrons in the narrow bands near the Fermi energy become localized,
and the system becomes a Mott insulator\cite{PhysRevB.75.193103,RC1977297}.

In past decades, the nature of the disorder-driven metal-insulator transition in two dimensional (2D) interacting system
has been discussed intensively\cite{A.M.Finkelstein,finkel1984weak,PhysRevB.30.527,PhysRevB.57.R9381,PhysRevLett.88.016802,doi:10.1126/science.1115660,app9061169}.
The existence of a metal state at zero magnetic filed was firstly predicted by Finkelstein\cite{A.M.Finkelstein,finkel1984weak} and Castellani et al\cite{PhysRevB.30.527}, and the possibility of metallic behavior and metal-insulator transition were later confirmed in Refs.\cite{PhysRevLett.88.016802,PhysRevB.57.R9381}.
By perturbative renormalization group methods, the combined effects of interactions and disorder were studied, and a quantum critical point was identified to separate the metallic phase stabilized by electronic correlation from insulating phase where disorder prevails over the electronic interactions\cite{doi:10.1126/science.1115660}. For reviews, see Refs.\cite{app9061169} and references therein. To understand the metal-insulator transition, it is now believed that we must consider
both electronic correlation and disorder on the same footing
because disorder and interactions are both present in real
materials\cite{Curro_2009,Dagotto257,PhysRevB.50.8039}. From a theoretical point of view, this is difficult.
When both disorder and interactions are strong, perturbative approaches usually break down\cite{RevModPhys.66.261,PhysRevB.30.527},
and quantum Monte Carlo simulations may be affected by the `minus-sign problem'.

In the context of QMC simulations, various interesting metal-insulator transitions
have been reported in different physical systems\cite{PhysRevLett.120.116601,PhysRevLett.109.026404,PhysRevB.101.155413}. By studying the disordered
Hubbard model on a square lattice at quarter filling, it was shown that repulsion between electrons can significantly enhance the conductivity, which provides evidence of a phase transition, in a two-dimensional model containing both interactions and disorder\cite{PhysRevLett.83.4610}.
The effects of a Zeeman magnetic field on the transport and thermodynamic properties have also been discussed\cite{PhysRevLett.90.246401};
it was argued that a magnetic field enhances localized behavior in the presence of interactions and disorder and induces a metal-insulator transition, in which the qualitative features of magnetoconductance agree with experimental findings.
In a two-dimensional system of a honeycomb lattice that
features a linearly vanishing density of states at the Fermi level, a novel disorder-induced nonmagnetic
insulating phase is found to emerge from the zero-temperature quantum critical point, separating a
semimetal from a Mott insulator\cite{Singha1176}.
The authenticity of the insulating phase has also been studied, and `false insulating'  behavior originates in closed-shell effects\cite{PhysRevB.85.125127}.

However, due to the limitation of the `minus-sign problem' in QMC simulations, most studies have focused on the half-filled case\cite{PhysRevLett.101.086401,PhysRevB.81.075106} or some fixed electronic filling\cite{PhysRevB.77.075101,PhysRevB.67.205112}.
Experimentally, transport measurements of effectively two-dimensional (2D) electron systems in
silicon metal-oxide-semiconductor field-effect transistors
provided evidence that a metal-insulator transition can occur,
where the temperature dependence of the conductivity $\sigma_{dc}$ changes
from that typical of an insulator at lower density to that typical of a conductor as the density increases
above a critical density\cite{PhysRevB.50.8039,PhysRevB.51.7038,PhysRevLett.77.4938}.
In two-dimensional Mott insulator also observed a transition from an anomalous metal to a Fermi liquid by doping\cite{doi:10.1126/science.abe7165}.
Thus, doping is also an important physical parameter to tune the phase transition,
while determining the doping-dependent metal-insulator transition is a subtle and largely understudied problem.
And study reported that cold atom-based quantum simulations offer remarkable opportunity for investigate the doping problem\cite{doi:10.1126/science.aal3837}.

In this paper, we evaluate the doping-dependent sign problem and then select several doping levels to examine the doping-dependent metal-insulator transition of the disordered Hubbard model on a square lattice. We then
examine whether this model also has a universal value of conductivity.
In simulations, the sign problem is minimized by
choosing off-diagonal values rather than diagonal disorder because, at
least at half filling, there is no sign problem in the
former case, and consequently, simulations can be pushed
to significantly lower temperatures.
We show that the sign-problem behavior worsens with increasing parameter strength, such as on-site interaction; however, the sign-problem behavior also decreases in the presence of bond disorder\cite{PhysRevB.55.4149}.
For results away from half filling, we choose some points, where the sign problem is
less severe compared to other densities, and show a phase diagram of the critical disorder strength determined by repulsion and doping in a disordered Hubbard model, going beyond previous results\cite{PhysRevLett.83.4610}.

\section{Model and method}
\label{sec:model}

The Hamiltonian for a disordered Hubbard model on a square lattice is defined as
\begin{eqnarray}
\label{Hamiltonian}
\hat H=-\sum_{{\bf ij}\sigma}t_{\bf ij}\hat c_{{\bf i}\sigma}^\dagger \hat c_{{\bf j} \sigma}^{\phantom{\dagger}}+U\sum_{{\bf i}}\hat n_{{\bf i}\uparrow}\hat n_{{\bf i}\downarrow}-\mu \sum_{{\bf i}\sigma} \hat n_{{\bf i}\sigma}
\end{eqnarray}
where $t_{\bf ij}$ and $U$ represent the hopping amplitude between the nearest-neighbor electrons
and on-site repulsive interaction,respectively, and $\mu$ denotes the chemical potential, which can control the electron density of the system.
$\hat c_{{\bf i}\sigma}^\dagger(\hat c_{{\bf i}\sigma}^{\phantom{\dagger}})$ is the creation (annihilation)
operator with spin $\sigma$ at site ${\bf i}$, and
$\hat n_{{\bf i}\sigma}$=$\hat c_{{\bf i}\sigma}^\dagger \hat c_{{\bf i}\sigma}^{\phantom{\dagger}}$ is the number operator.
Disorder is introduced by taking the hopping parameters $t_{\bf ij}$ from a probability $P(t_{\bf ij})=1/\Delta$
for $t_{\bf ij}\in[t-\Delta/2,t+\Delta/2]$ and zero otherwise. $\Delta$ is a measure of the strength of the disorder\cite{PhysRevLett.83.4610}. We set $t$=1 as the default energy scale.
The number of disorder realizations used in present work is $20$ which is enough to obtain reliable results (see Appendix for details).

We use the DQMC method\cite{PhysRevD.24.2278} to investigate the phase transitions in the model defined by Eq.(\ref{Hamiltonian}) numerically. DQMC is a nonperturbative approach, providing an exact numerical method to study
the Hubbard model under a finite temperature. First, the partition function $Z=Tr e^{-\beta H}$ is regarded as a path integral discretized
into $\Delta \tau $ functions in the imaginary time interval $(0,\beta)$.
The kinetic term is quadratic, and the on-site interaction term can be decoupled into a quadratic term by a discrete Hubbard-Stratonovich field; then, by analytically integrating the Hamiltonian quadratic term, $Z$ can be converted into the product of two fermion determinants, where one is spin up and the other is spin down.
The Metropolis algorithm is used to stochastically update the sample, and we set $\Delta\tau=0.1$, leading to sufficiently small errors in the Trotter approximation.

To study the phase transitions of the system, we computed the $T$-dependent dc conductivity, which can be obtained from the momentum $\textbf{q-}$ and imaginary time $\tau$-dependent current-current correlation function $\Lambda_{xx}(\textbf{q},\tau)$\cite{PhysRevB.54.R3756,PhysRevLett.75.312}:
\begin{eqnarray}
\label{DC}
\sigma_{dc}(T)=\frac{\beta^2}{\pi}\Lambda_{xx}(\textbf{q}=0,\tau=\frac{\beta}{2})
\end{eqnarray}
Here, $\Lambda_{xx}(\textbf{q},\tau)$=$\left<\hat{j}_x(\textbf{q},\tau)\hat{j}_x(\textbf{-q},0)\right>$, $\beta$=$1/T$, where $\hat{j}_x(\textbf{q},\tau)$ is the Fourier transform of time-dependent current operator $\hat{j}_x(\textbf{r},\tau)$ in the $x$ direction:
\begin{eqnarray}
\label{transform}
\hat{j}_x(\textbf{r},\tau) = e^{H\tau/h}\hat{j}_x(\textbf{r})e^{-H\tau/h}
\end{eqnarray}
where $\hat{j}_x(\textbf{r})$ is the electronic current density operator, defined in Eq.(\ref{J}).
\begin{eqnarray}
\label{J}
\hat{j}_x(\textbf{r}) = {i}\sum_{\sigma}t_{i+\hat{x},i}\times(c_{i+\hat{x},\sigma}^{+}c_{i\sigma}-c_{i \sigma}^{+}c_{i+\hat{x},\sigma})
\end{eqnarray}
The validity of Eq.(\ref{DC}) has been examined, and this equation has been used for metal-insulator transitions
in the Hubbard model in many studies\cite{PhysRevLett.75.312,PhysRevLett.83.4610,PhysRevLett.120.116601}.

\section{Results and discussion}

\label{sec:results}

\begin{figure}[t]
\centerline {\includegraphics[width=3.2in]{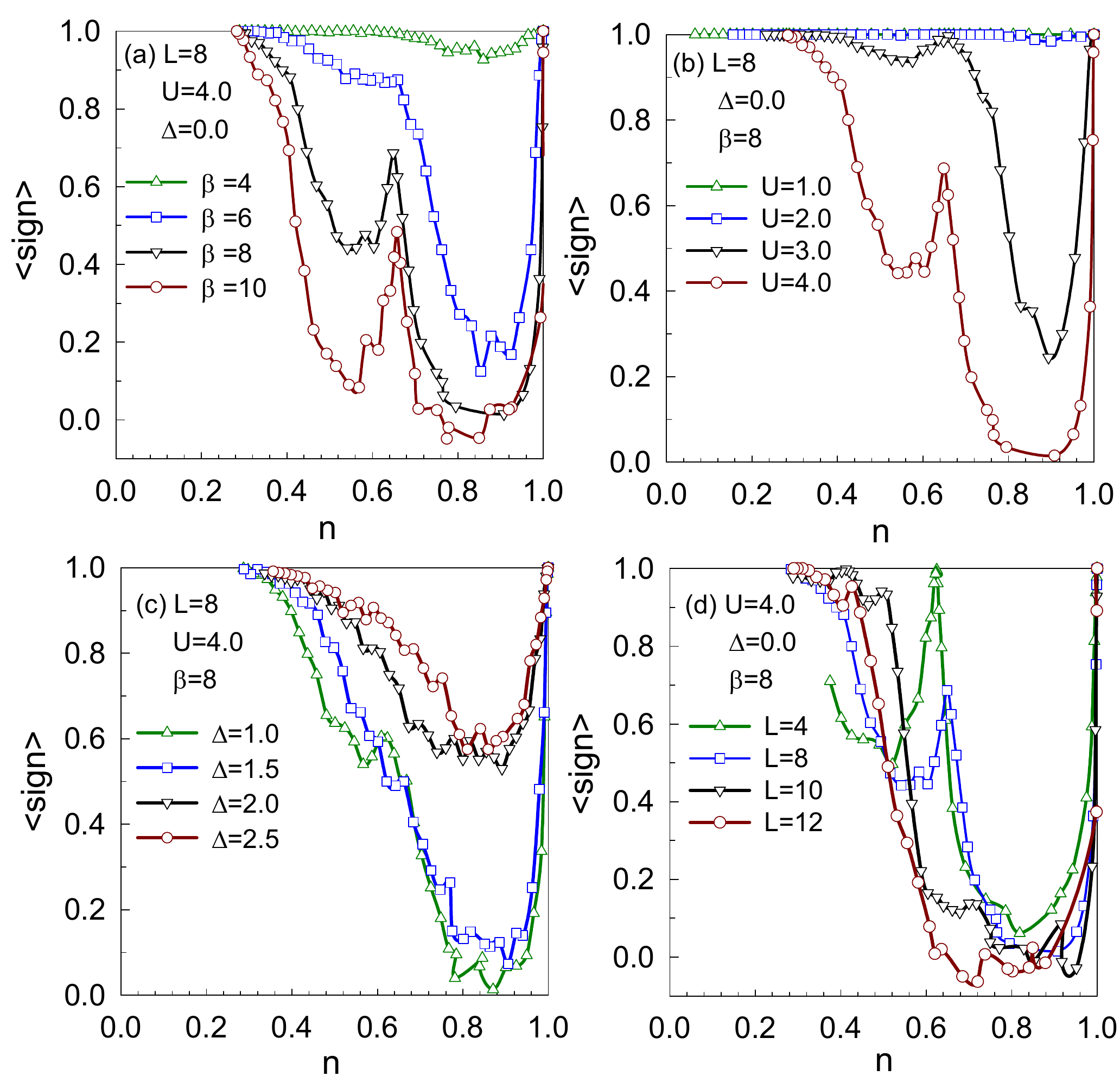}}
\caption{(Color online) Average sign as a function of electron filling for (a) different temperatures, (b) different interactions, (c) different disorder strengths, and (d) different lattice sizes.}
\label{Fig:sign}
\end{figure}

At half filling, due to the particle-hole symmetry, under the transformation $c_{\bf{i}\sigma}^{\dagger} \rightarrow (-1)^{\bf{i}}c_{\bf {i}\sigma}$,
the Hamiltonian is unchanged, and
the simulation can be performed without considering the sign problem\cite{PhysRevLett.87.146401}.
When far from half filling, the system may have a sign problem; thus, in a doped Hubbard model on a square lattice, the notorious sign problem prevents exact results at lower temperatures, at higher interactions, or with larger lattices.
To ensure the reliability of the data in our simulation,
we first present the average sign in Fig.\ref{Fig:sign},
which is shown as a function of electron filling for (a) different temperatures, (b) different interactions, (c) different disorder strengths, and (d) different lattice sizes along with the Monte Carlo parameters after 30,000 iterations.
The average sign decays exponentially both with increasing inverse temperature and lattice size\cite{PhysRevB.55.4149}.

The average sign is determined by the ratio of the integral of the product of up and down spin determinants to the integral of the absolute
value of the product\cite{PhysRevB.92.045110}:
\begin{eqnarray}
\label{J}
\langle S \rangle &=
\frac
{\sum_{\cal X} \,\,
{\rm det} M_\uparrow({\cal X}) \,
{\rm det} M_\downarrow({\cal X})
}
{
\sum_{\cal X}  \,\,
| \, {\rm det} M_\uparrow({\cal X}) \,
{\rm det} M_\downarrow({\cal X}) \, |
}
\end{eqnarray}
where ${\cal X}$ is the HS configurations composed of the spatial sites and the imaginary time slices;
and $M_\sigma({\cal X})$ is defined as each spin specie matrix.
As shown in Fig.\ref{Fig:sign} (a), we evaluate the variation in the sign problem with density for various inverse temperatures.
The average sign decreases quickly as the system is doped from $n=1.0$ to $n=0.9$.
The average sign is small when $0.68<n<0.98$, with a value below 0.2 due to the disappearing signal-to-noise ratio in the data,
making DQMC simulations nearly impossible.
As $n$ decreases from 0.68, the average sign increases and then decreases from $n$=0.64 until
$n=0.56$, after which the average sign continuously increases with decreasing density.
Thus, the sign problem is acceptable only at some specific densities, which is correlated with the closed-shell effects.
Moreover, comparing various temperatures, the sign problem becomes worse as $T$ decreases.
Fig.\ref{Fig:sign} (b) shows the effect of the on-site interaction on the sign problem and indicates that the sign problem is more serious with increasing interaction; in other words,
the interaction plays a negative role in the average sign.
The influence of the bond disorder on the sign problem is shown in Fig.\ref{Fig:sign} (c). By raising the disorder strength, the sign problem improves,
fundamentally differing from local site disorder, which breaks the particle-hole symmetry\cite{PhysRevLett.98.046403} and enhances the sign problem.
Fig.\ref{Fig:sign} (d) shows that lattice size also affects the sign problem, and the average sign is smaller for $L=10,12$ than for $L=8$.

\begin{figure}[t]
\centerline {\includegraphics*[width=3.6in]{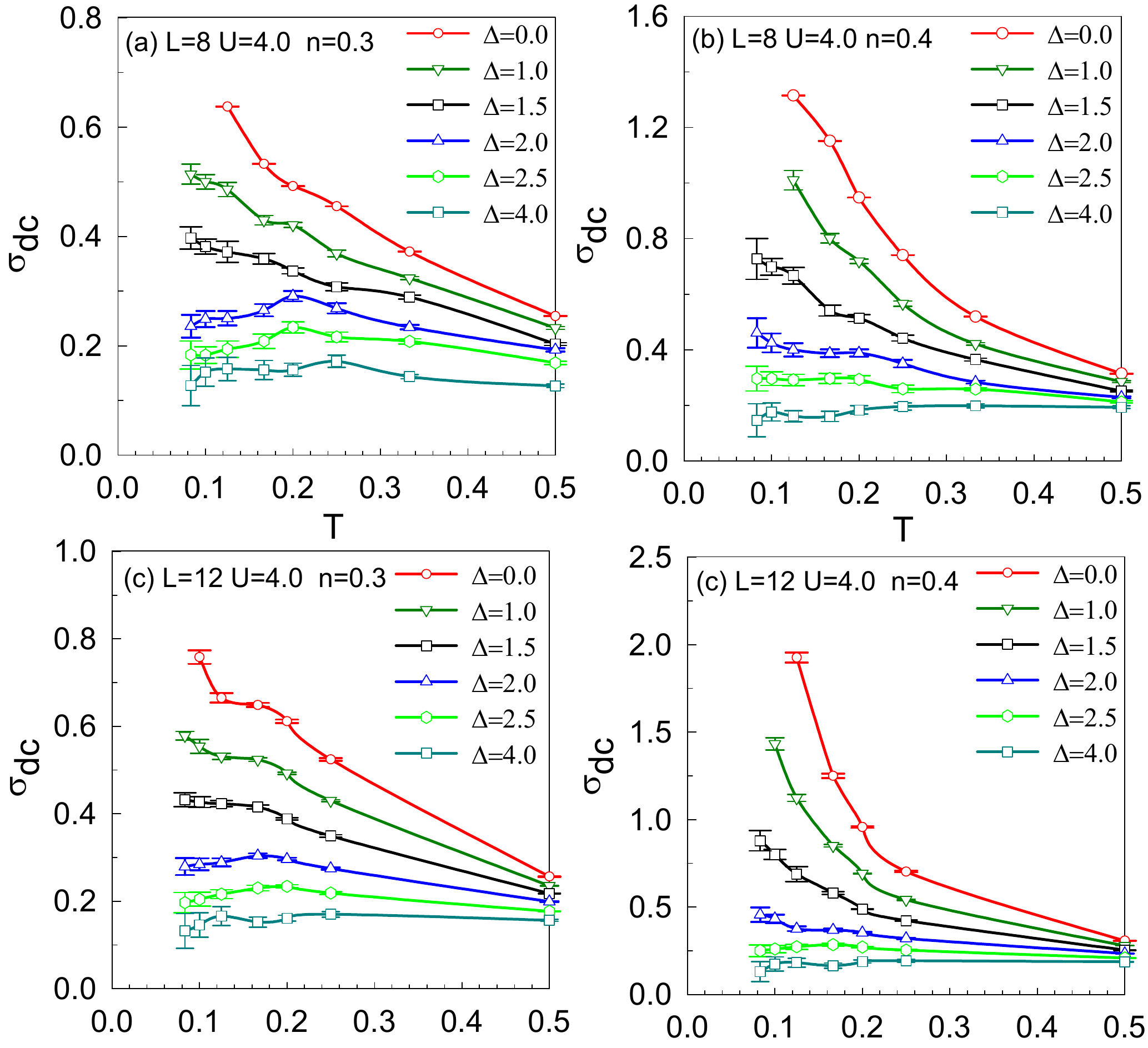}}
\caption{(Color online) Conductivity $\sigma_{dc}$ as a function of temperature
at $U=4.0$ for different $\Delta$ at (a) and (c) $n=0.3$; (b) and (d) $0.4$. Top panel is about $L=8$, and lower panel is about $L=12$.}
\label{Fig:UD}
\end{figure}

\begin{figure}[t]
\centerline {\includegraphics*[width=3.5in]{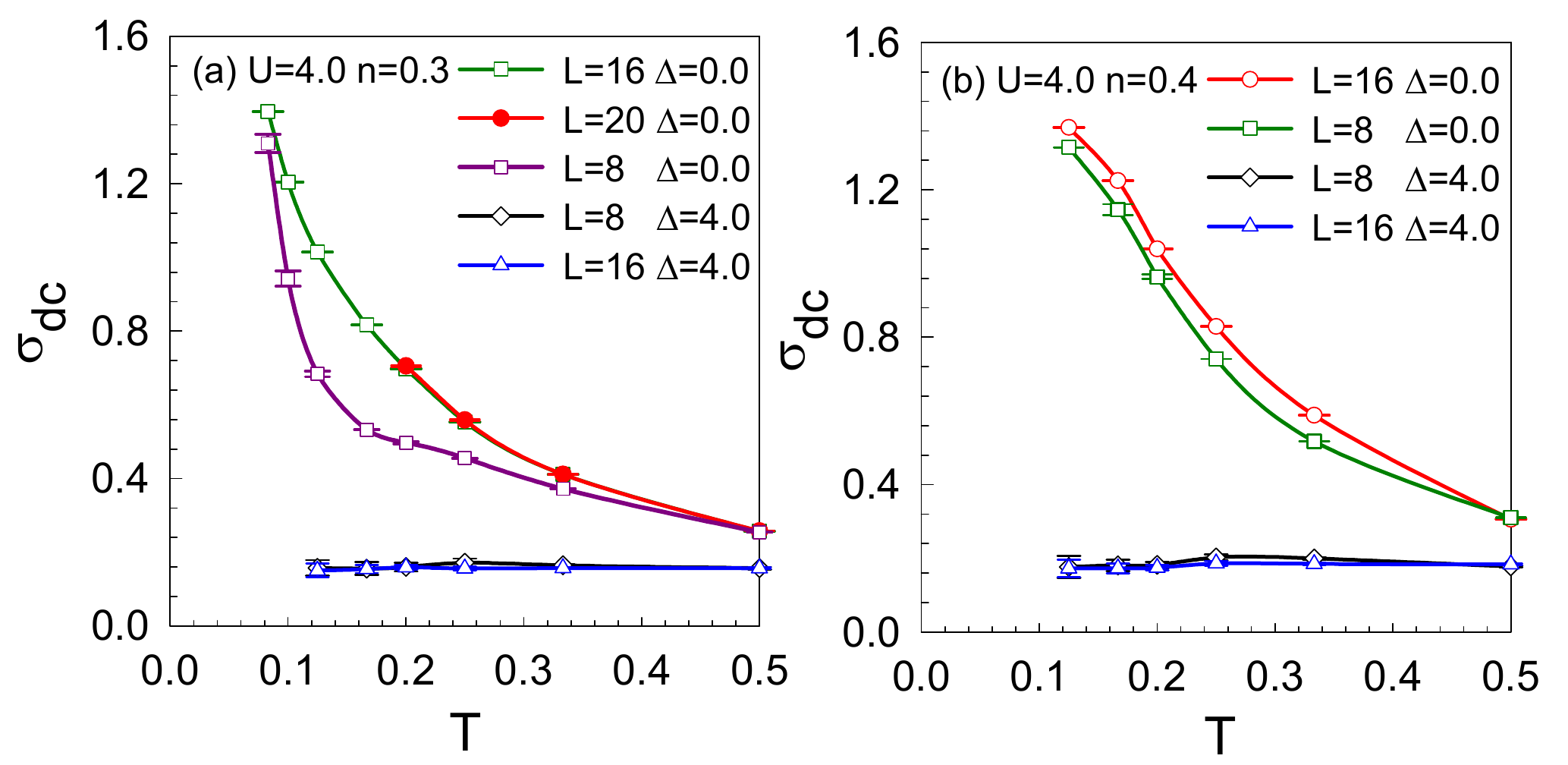}}
\caption{(Color online) Conductivity $\sigma_{dc}$ as a function of temperature for different disorder strengths for $U=4.0$ on the $L=8,16,20$ lattices at (a) $n=0.3$ and (b) $n=0.4$.
In (a), the $\sigma_{dc}$ curves of $L=20$ and $L=16$ are almost coincident which indicates that the dc conductivity tends to converge as $L\geq 20$, although the lower temperature calculation at $\beta >6$ is constrained by the DQMC simulations.}
\label{Fig:L}
\end{figure}

Based on these results, longer runs are required to make the results more reliable.
We thus perform simulations where $L=8$ and $U=4.0$, in which the sign problem is mild for the DQMC method and does not prohibit obtaining accurate results.
In general, in the absence of disorder and frustration, the ground state of the square lattice at half filling is sensitive to interactions, and the system becomes an AF insulator for any finite value of interaction $U>0$ due to the perfect nesting in the Fermi surface.
Previous studies demonstrated that when considering disorder at half filling for $U=4.0$, the insulating behavior at low temperatures persists to much larger bond disorder strengths\cite{PhysRevLett.83.4610}.
According to previous studies, a basic question arises:
On a square lattice with repulsive interactions, in addition to half filling, how are the transport properties at other carrier concentrations affected by disorder?
To answer this question, we take advantage of the temperature-dependent dc conductivity $\sigma_{dc}(T)$ to distinguish between an insulator and a metal.
Fig.\ref{Fig:UD} shows $\sigma_{dc}(T)$ measured
on the square lattice across several disorder $\Delta$ values at different densities $n=0.3,0.4$,
where the sign problem has little effect on the results.
In the low-temperature regime, the behavior of $\sigma_{dc}$ shows that a transition from metallic to insulating behavior occurs with increasing disorder.
For example, when $L=8$, $n=0.3$, $T\leqslant0.2$ and $\Delta=0.0$, the dc conductivity grows as the temperature decreases (i.e., $d\sigma_{dc}/dT<0$), which indicates that the system is metallic, and the error bars stem from the statistical fluctuation of disorder sampling.
Conversely, at $\Delta=4.0$, the dc conductivity falls with decreasing temperature (i.e.,
$d\sigma_{dc}/dT>0$) and approaches zero as $T\rightarrow0$, which is characteristic of insulating behavior.
Therefore, it can be deduced from the above figure that hopping disorder decreases the dc conductivity.
The transition from metallic to insulating clearly occurs at $\Delta_{c}=1.5\thicksim2.0$.
In the same way, by changing the carrier density $n$, the critical disorder strength
at $n=0.4$ is about $\Delta_{c}=2.5$,
indicating the occurrence of the metal-insulator transition in the presence of disorder at other densities, which differs from the half filling case.
Fig.\ref{Fig:UD}(c), (d) show the results of $L=12$. Even though the values of dc conductivity have not been saturated at $L=12$, the values of critical disorder strength are roughly the same for $L=8$ and $L=12$. Our further data in Fig. \ref{Fig:L} (a) show that the dc conductivity itself tends to converge at $L=20$, while simulations on such lattice cost huge CPU times.
And through the shift in the maximum dc conductivity, one can infer that the mobility gap increases as the bond disorder increases.

In addition, we ascertain that the occurrence of a phase transition results from the bond disorder rather than the system size being smaller than a localization length.
Fig.\ref{Fig:L} (a) shows that as the lattice size increases, the dc conductivity will converge
to a finite value under various conditions, although the convergence speed is affected by parameter sets (such as $\sigma_{dc}$ in the insulating phase converges faster than in the metallic phase, or $\sigma_{dc}$ in system at $n=0.4$ converges faster than $n=0.3$).

\begin{figure}[t]
\centerline {\includegraphics*[width=3.2in]{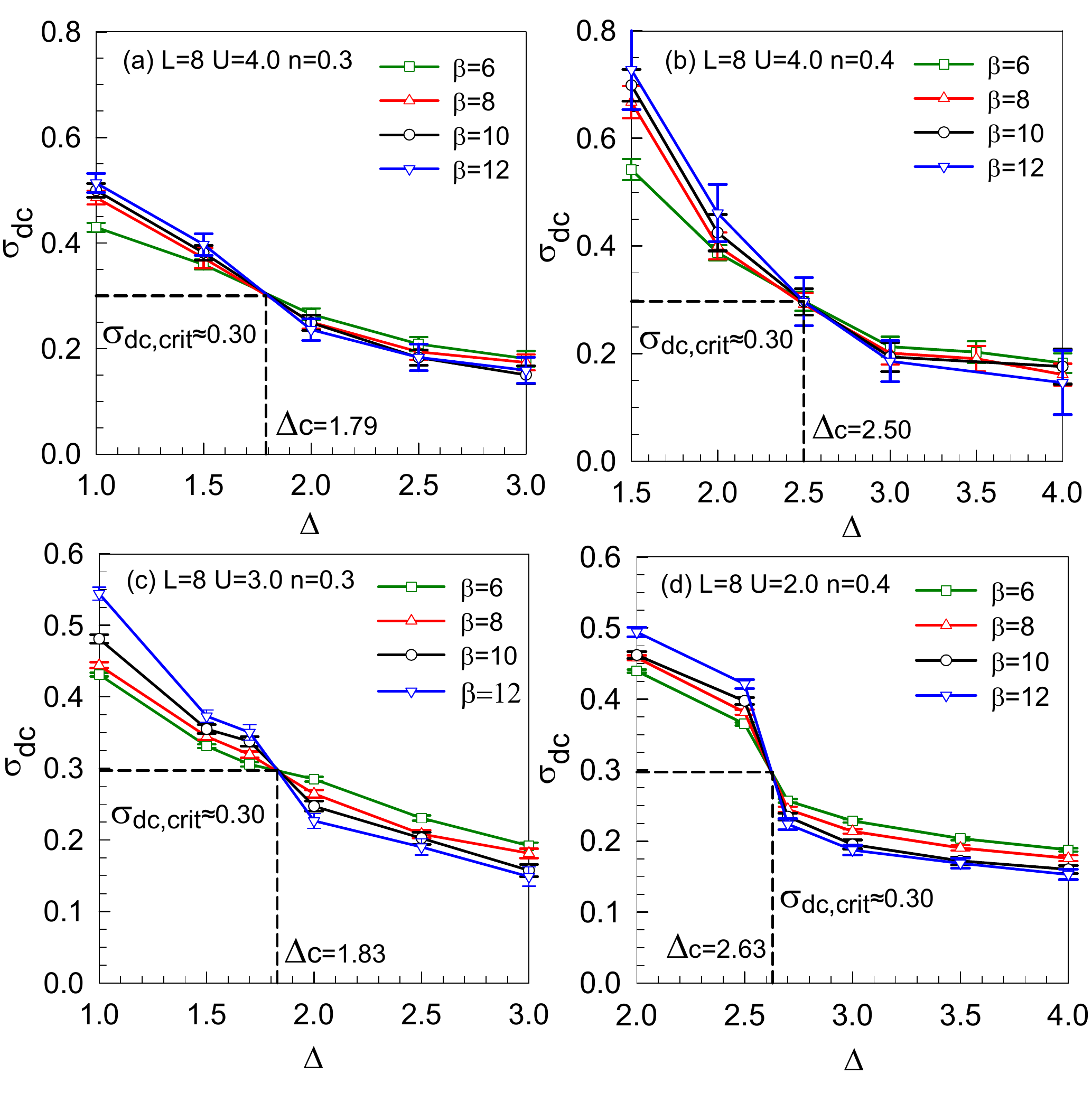}}
\caption{(Color online) Conductivity as a function of the disorder strength for three temperature $\beta=6,8,10,12$
at (a) $U=4, n=0.3$; (b) $U=4, n=0.4$ (c) $U=3, n=0.3$; and (d) $U=2, n=0.4$. The intersection determines the critical disorder strength, and the value of the conductivity at the critical disorder is approximately 0.30.}
\label{Fig:universal}
\end{figure}

\begin{figure}[t]
\centerline {\includegraphics*[width=3.2in]{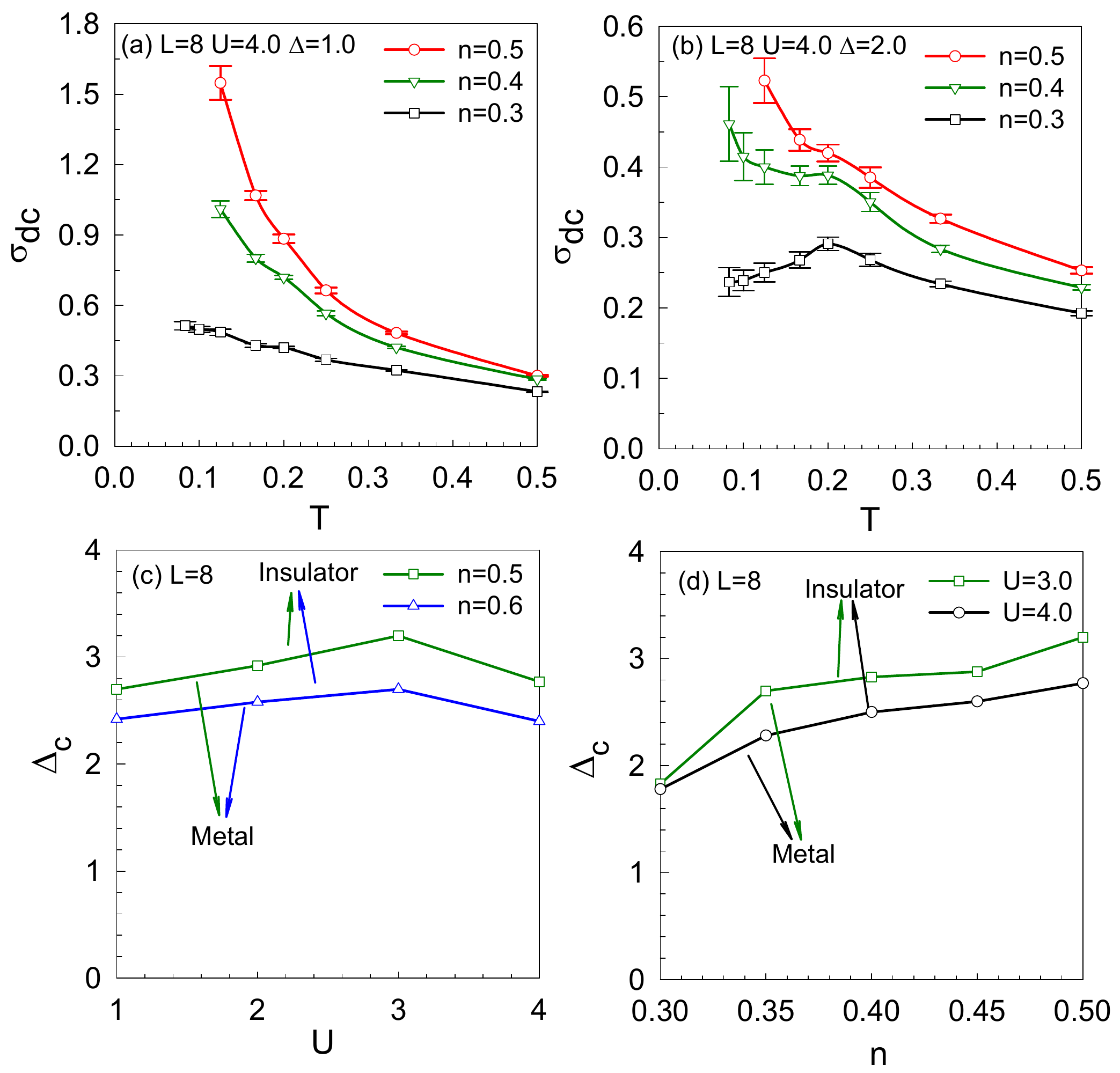}}
\caption{(Color online)Top: conductivity $\sigma_{dc}$ as a function of temperature at $U=4.0$ for (a) $\Delta$=1.0 and (b) $\Delta$=2.0 with different fillings. Below: critical disorder strength $\Delta_{c}$ (c) as a function of $U$ at different $n$ and (d) as a function of $n$ at different $U$.}
\label{Fig:U}
\end{figure}

\begin{figure}[t]
\centerline {\includegraphics*[width=3.2in]{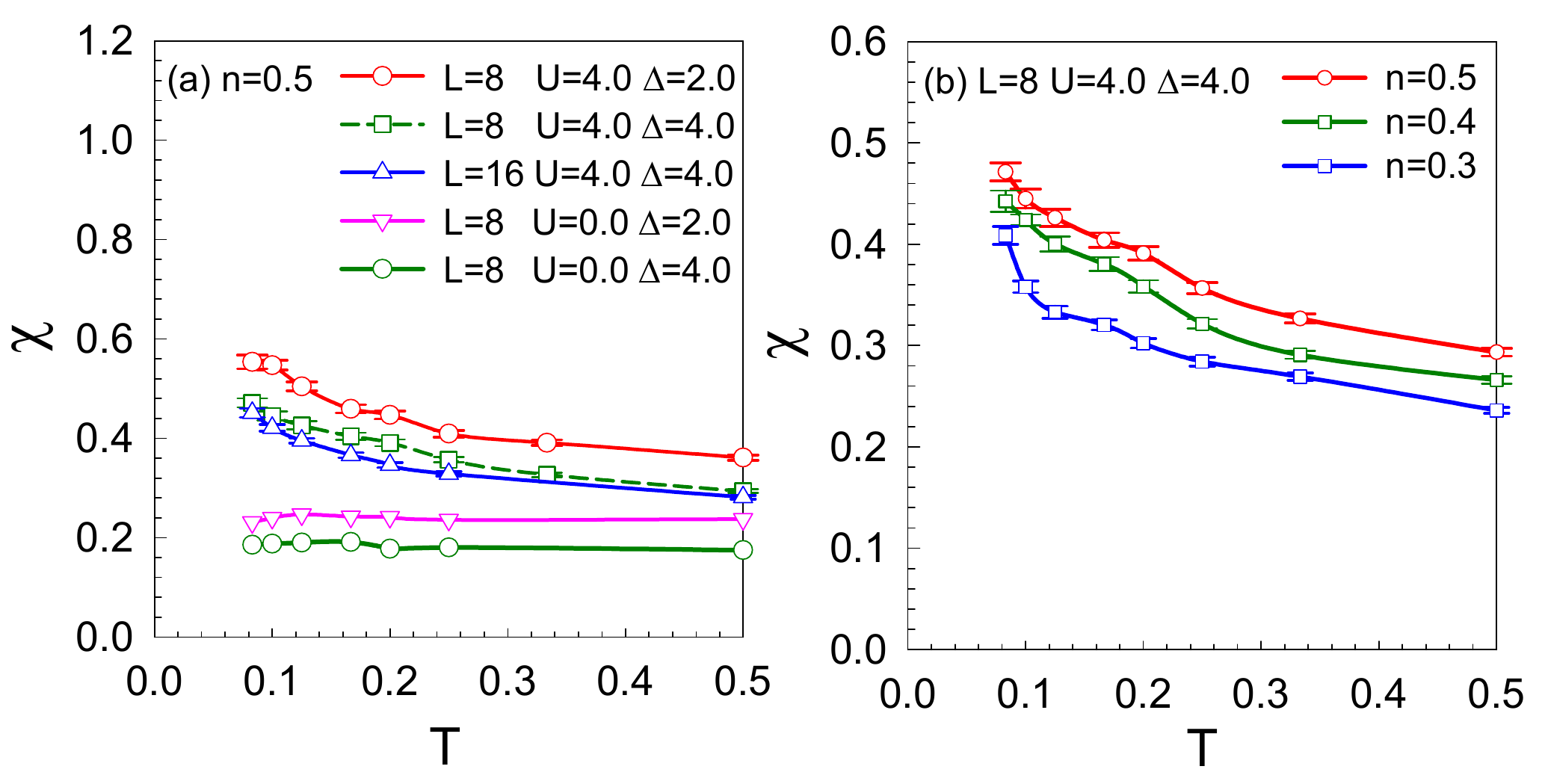}}
\caption{(Color online) Spin susceptibility $\chi$ as a function of temperature (a) at various interaction strengths $U=0.0,4.0$, disorder strengths $\Delta=2.0,4.0$ and lattice sizes $L=8,16$ at fixed density $n=0.5$ and (b) at fixed interaction strength $U=4.0$ and disorder strength $\Delta=4.0$ with different fillings on an $N=8\times 8$ square lattice.}
\label{Fig:X}
\end{figure}

On the basis of Fig.\ref{Fig:UD}, we plot $\sigma_{dc}$ as a function of the disorder strength in Fig.\ref{Fig:universal}, to determine the critical point accurately and the corresponding value of dc conductivity.
The intersection of four curves marks the critical point for the metal-insulator transition.
The ordinate of this intersection describes the critical dc conductivity (i.e.,
at $n= 0.3$, $\sigma_{dc,crit}=0.30$, and at $n= 0.4$, $\sigma_{dc,crit}=0.30$). Here, the value of the critical dc conductivity is determined to an accuracy of 0.01. 
Comparing the results for these parameters sets ($U=4.0, n=0.3$ and $U=4.0, n=0.4$), it shows that the system has the possibility of a universal value of the critical dc conductivity.
To strongly support these findings, we present the same plots for different interaction strength ($U=2.0,3.0$) shown in Fig.\ref{Fig:universal} (c) and (d): although the critical disorder strength is varied,
the critical dc conductivity is still $\sigma_{dc,crit}=0.30$.
Besides, we also compute other parameter sets, such as $U=4.0$, $n= 0.5$, $\Delta_{c}=2.77$, $\sigma_{dc,crit}=0.30$; $U=3.0$, $n= 0.6$, $\Delta_{c}=2.70$, $\sigma_{dc,crit}=0.26$;
$U=2.0$, $n= 0.5$, $\Delta_{c}=2.91$, $\sigma_{dc,crit}=0.29$; and $U=1.0$, $n= 0.6$, $\Delta_{c}=2.42$, $\sigma_{dc,crit}=0.32$. The standard deviation equal to $0.02$ is small enough to ensure the clustering of the dc conductivity values around the mean value,
which confirms the existence of universal conductivity($\sigma_{dc,crit}=0.30\pm0.01$)\cite{PhysRevB.84.035121} (the error 0.01 is computed by estimating the arithmetic mean from the listed eight datasets) and its independence with $n$, $U$, and $\Delta_{c}$. 
This property has also been realized in the quantum sigma model\cite{Anissimova2007,Punnoose289}, and discussed in both
graphene\cite{PhysRevLett.98.256801} and integer quantum Hall effect\cite{PhysRevLett.95.256805}.

To describe the role of doping in more details, we investigate the change in $\sigma_{dc}$
with different densities at fixed disorder strength,
as shown in Fig.\ref{Fig:U} (a) and (b). Increasing the electronic density shall enhance the dc conductivity, and when $\Delta=2.0$, the system behaves as an insulator at $n=0.3$.
Conversely, at $n=0.4$ and $n=0.5$, the system behaves as a metal.
Thus, we deduce that doping can affect the metal-insulator transition.
We compile the results of $\Delta_{c}$ in Fig.\ref{Fig:U} (c), (d), showing the relationship between critical disorder strength and interaction strength $U$ (or density $n$).
The critical disorder strength increases firstly and then decreases as $U$ increases at a fixed density,
which is also reported in the ionic Hubbard model\cite{PhysRevLett.98.046403,PhysRevB.99.014204}.
The Coulomb repulsion enhances metallicity when $U<3.0$, and a larger $U$ will make it more effective to localize electrons to decrease $\sigma_{dc}$.
On the other hand, in our calculation, the effect of density on $\Delta_{c}$ is non-monotonous:
$\Delta_{c}$ increases as the density increases from 0.3 to 0.5, and then decreases as the density increases to 0.6.
Although the sign problem restricts us to calculate the large density, the current results have provided strong support for the conclusion that doping-dependent metal-insulator transition in a disordered Hubbard model.

The spin dynamics of electrons are often discussed together with the localization transition, and we discuss the correlation between the spin susceptibility and temperature through $\chi=\beta S(q=0)$, where $S(q=0)$ denotes the ferromagnetic structure factor\cite{PhysRevB.59.3321}.
Fig.\ref{Fig:X} (a) shows that the spin susceptibility $\chi$ increases as the temperature decreases and as $U$ increases (for $U=0.0$ and $U=4.0$),
meaning that interaction can enhance the ferromagnetic susceptibility. Additionally, the spin susceptibility
diverges as $T\rightarrow0$, implying that magnetic order exists in both the metallic ($\Delta=2.0$) and insulating phases ($\Delta=4.0$).
The ferromagnetic susceptibility reduces with increasing disorder in the presence of interaction and hopping disorder, which is in accord with the Stoner criterion for a ferromagnetic $UN(E_{F})>1$.
$N(E_{F})$ represents the density of states at the Fermi level.
The Stoner criterion estimates that the behavior of a ferromagnetic acts against increasing disorder due to the reduction in the spectral density at the Fermi level\cite{PhysRevB.84.155123}.
Comparing the results of $L=8$ with $L=16$, the spin susceptibility is little affected by size effects.
Additionally, we find that the density plays a positive role in the ferromagnetic susceptibility, as shown in Fig.\ref{Fig:X} (b).
\section{Conclusions}
\label{sec:conclusions}

In summary, we have studied a disordered Hubbard model on a square lattice away from half filling by using the determinant quantum Monte Carlo method. We find that the sign problem emerges away from half filling, accompanied by a nonmonotonic behavior as the density varies, and that adding hopping disorder can reduce the sign problem.
The system becomes metallic at finite $U$ unlike with half filling, and the metal-insulator transition is affected by disorder.
Although the critical disorder strength non-monotonically varies with changing the electron density and repulsion,
the critical dc conductivity is independent of the parameter set, similar to the site disorder case\cite{PhysRevB.84.035121}.
The behavior of spin susceptibility suggests that under a range of densities, the insulating phase is accompanied by local moments.
The ferromagnetic susceptibility tends to reduce with increasing bond disorder strength, in line with the Stoner criterion.

At fixed disorder, we also demonstrate that the carrier density $n$ can be used as a tuning parameter for the occurrence of the phase transition, which can be explained as follows:
varying the intensity of disorder $\Delta$ at a fixed density $n$ can be regarded as adjusting the mobility boundary via the Fermi energy and is similar to varying the carrier concentration $n$ at a fixed disorder strength $\Delta$, which can be thought of as a shift in the Fermi energy\cite{PhysRevLett.83.4610}.

\begin{acknowledgments}
This work is supported by NSFC (Nos. 11974049 and 11774033) and Beijing Natural Science Foundation (No. 1192011). The numerical simulations were performed at the HSCC of Beijing Normal University and on the Tianhe-2JK in the Beijing Computational Science Research Center.
\end{acknowledgments}

\begin{figure}[htb]
\centerline {\includegraphics*[width=3in]{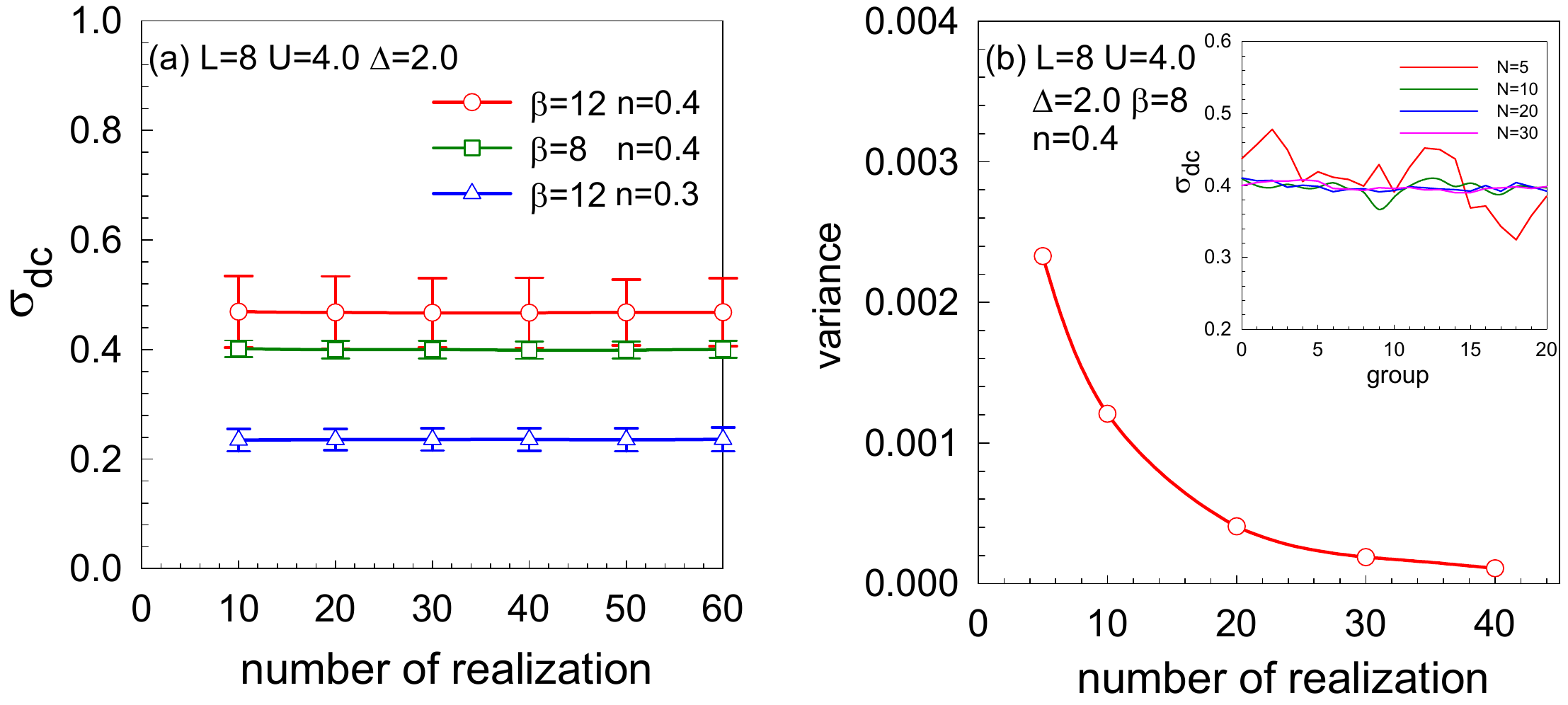}}
\caption{(Color online)(a)Conductivity $\sigma_{dc}$ as a function of number of realization at $L=8, U=4.0$, and $\Delta=2.0$. The error bars are derive from DQMC simulation. (b) The corresponding variance of the data in the insert.
Insert: The averaged dc conductivity as a function of the number of groups. N represents the number of disorder realizations in a group.}
\label{Fig:number}
\end{figure}

\appendix

\setcounter{equation}{0}
\setcounter{figure}{0}
\renewcommand{\theequation}{A\arabic{equation}}
\renewcommand{\thefigure}{A\arabic{figure}}
\renewcommand{\thesubsection}{A\arabic{subsection}}

\section{Concerning the number of disorder realizations}

In general, the required number of disorder realizations must be determined empirically, and is a complex interplay between ``self-averaging" on sufficiently large lattices, the strength of the disorder,
and the location in the phase diagram.
In Fig.\ref{Fig:number}, we show the change of the average dc conductivity with the number of random disorder realizations.
For any given density $n$, no change in the average $\sigma_{dc}$ for realization numbers larger than 10.
It justifies our usage of $20$ realizations in the main text.

We also use the variance to justify our choice of the number of disorder realizations.
In the inset of Fig.\ref{Fig:number} (b), we calculated the average values of several groups of data whose realizations are 5, 10, 20, 30 respectively, and performed 20 times. It can be seen that the average values of each group with N=5 vary greatly and the curve fluctuates violently, means that N=5 can not eliminate the random error well. When the number in a group increased to be 10, fluctuations were significantly suppressed; increased to be 20, the curve tends to be stable. This phenomenon is also shown in Fig.\ref{Fig:number} (b) in the form of variance of each curve. The variance curve shows good convergence. As the number in a group increases to be 20, the variance has decreased to a value close to 0. That is, 20 times is large enough as the number of realization.

\bibliography{reference}

\end{document}